\documentclass[jcp,twocolumn,showpacs,preprintnumbers]{revtex4}%
\usepackage{graphics}
\usepackage{bm}
\usepackage{amsmath}
\usepackage{amsfonts}
\usepackage{amssymb}
\usepackage{subfig}
\usepackage{graphicx}%
\usepackage{color}
\begin{document}

\title{A study for the static properties of symmetric linear multiblock copolymers under poor solvent conditions}

\author{Panagiotis E. Theodorakis}

\affiliation{Faculty of Physics, University of Vienna,
Botlzmanngasse 5, A-1090 Vienna, Austria}

\affiliation{Institute for Theoretical Physics and Center for
Computational Materials Science, Vienna University of Technology,
Hauptstra{\ss}e 8-10, A-1040 Vienna, Austria}

\affiliation{Vienna Computational Materials Laboratory,
Sensengasse 8/12, A-1090 Vienna, Austria}

\author{Nikolaos G. Fytas}
\affiliation{Departamento de F\'{i}sica
Te\'{o}rica I, Universidad Complutense, E-28040 Madrid, Spain}

\date{\today }

\begin{abstract}
We use a standard bead-spring model and molecular dynamics
simulations to study the static properties of symmetric linear
multiblock copolymer chains and their blocks under poor solvent
conditions in a dilute solution from the regime close to theta
conditions, where the chains adopt a coil-like formation, to the
poorer solvent regime where the chains collapse obtaining a
globular formation and phase separation between the blocks occurs.
We choose interaction parameters as is done for a standard model,
i.e., the Lennard-Jones fluid and we consider symmetric chains,
i.e., the multiblock copolymer consists of an even number $n$ of
alternating chemically different A and B blocks of the same length
$N_{A}=N_{B}=N$. We show how usual static properties of the
individual blocks and the whole multiblock chain can reflect the
phase behavior of such macromolecules. Also, how parameters, such
as the number of blocks $n$ can affect properties of the
individual blocks, when chains are in a poor solvent for a certain
range of $n$. A detailed discussion of the static properties of
these symmetric multiblock copolymers is also given. Our results
in combination with recent simulation results on the behavior of
multiblock copolymer chains provide a complete picture for the
behavior of these macromolecules under poor solvent conditions, at
least for this most symmetrical case. Due to the standard choice
of our parameters, our system can be used as a benchmark for
related models, which aim at capturing the basic aspects of the
behavior of various biological systems.
\end{abstract}


\maketitle

\section{INTRODUCTION}
\label{sec:1}

Polymers combining  chemically different monomeric units on their
structure found very much theoretical and experimental interest,
as, for instance, diblock copolymers, where in the case of a
linear chain one block of the chain consists of one type of
monomers, while the other part of another type of monomers. These
macromolecules have already found particular interest as they form
different structures depending on their structural parameters,
thermodynamics conditions, etc. For a standard theoretical review
one could see Ref.~\cite{1} and references therein. We underline
that the motivation for studying such systems stems from the fact
that these materials are used in industrial scale~\cite{1} and
also in very advanced technological applications (as an example
one may look a very recent review Ref.~\cite{2}). As an extension,
regarding possible complexity in the structure or other aspects,
of diblock copolymers is regarded multiblock copolymers, where in
their simplest case more than two blocks comprise a linear
polymeric chain and the blocks are composed either of type A or
type B monomers. The phase diagram of multiblock copolymer
melts~\cite{3} resembles the phase diagram of diblock copolymer
chains~\cite{4,5,6}, where the scaling parameter $\chi N$ controls
the phase behavior of the system, as it has also been discussed in
simulation studies~\cite{7}. For the most symmetrical case the
lamellar structure is predicted, an approach that is particular
valid in the strong segregation limit. The case of multiblock
copolymer chains is particularly interesting even when only two
type of blocks (A, B) alternating along a linear chain in an
infinite dilute solution is
considered~\cite{8a,8b,8c,8,9,10,11,12}. In this case, it is
sufficient to study the behavior of an isolated chain taking into
account only interactions (often of short range) within the chain.
A model describing a multiblock copolymer chain could also exhibit
some interest due to its close relation to various
toy-models~\cite{13}, which try to mimic the behavior of various
biomacromolecules in several processes~\cite{14}. Such
biomacromolecules are rather formed by periodically repeated
structural units (``monomers'') along their chain. Simple
(coarse-grained) models~\cite{15} incorporating  the key aspects
for a specific biomacromolecule are able to capture efficiently
their relevant properties for which the model is designed.

The phase behavior of a single multiblock copolymer chain has been
extensively studied by means of molecular dynamics (MD)
simulations by using a standard bead spring-model and considering
interactions between beads according to a standard model
(Lennard-Jones (LJ) fluid)~\cite{10,11,12}. Of course the use of
MD simulations imposes certain limits in the range that the
parameters may vary, for example, the range of the temperature and
the chain length. However, it has been shown that the different
regimes of interest for the phase behavior can all be accessed by
simulations~\cite{8a,8b,8c,10,11,12,15}. Multiblock copolymers
composed of two different types of blocks (A and B) alternating
along the linear chain and length of a block A being equal to the
length of block B that adopt coil-like structures in a good
solvent or at temperatures close to the
$\Theta$~\cite{8,9,10,11,12,16} have been studied. The chemical
difference of monomers should actually be kept responsible for an
expansion in the chain dimensions with respect to the equivalent
homopolymer chains (same total number of monomers) under the same
thermodynamic conditions~\cite{8,9}. Below a certain
temperature~\cite{10} a multiblock copolymer chain collapses due
to the incompatibility between monomers and the solvent obtaining
the so-called globule formation and different scenarios of phase
behavior have been discussed~\cite{11,12}. Simulations have
confirmed the validity of the parameter $\chi N$ in defining the
phase behavior of a di/multi-block copolymer system, but it has
also been discussed that the geometry of the microphase separated
regions is controlled by the total number of blocks $n$, as well
as other parameters, i.e. the relative size and arrangement of the
blocks~\cite{7}. The same has been shown to be true for a single
multiblock copolymer chain under poor solvent
conditions~\cite{10,11,12}.

The phase behavior of such macromolecules has been discussed
within the frame of the analysis of the formed clusters (which
contain two or more blocks of the same type) and of the
probability that a cluster of one type of blocks contains all A
blocks while another cluster contains all B
blocks~\cite{10,11,12}, i.e. a single lamellar structure with an
A-B interface between blocks of type A and blocks of type B. Thus,
when this probability is unity we fall into the above single
lamellar case. Then, for the range of temperatures ($\chi \propto
\Delta \varepsilon/T$, $\Delta \varepsilon = const$) and of chain lengths ($nN$,
where $N$ is the block length; blocks A and B have the same
length) accessible to simulations this type of phase behavior has
been seen for small number of blocks $n$ and for values of block
length $N>20$. Another scenario suggests that the above type of
phase behavior takes place with a certain probability
($P_{N_{cl}=2} < 1$, $N_{cl}$ being the number of clusters; such
probability can be very high or very small depending on the values
of $N$, $n$, and $T$). The third scenario corresponds to the case
that the probability $P_{N_{cl}=2} = 0$ (the formation of only two
clusters with monomers of different type never occurs) and a
symmetric variation in the number of clusters $N_{cl}$ around an
average value ($2<N_{cl}<n$, $n$ is the total number of blocks of
the chain) is seen. Of course, the analysis of the probability is
performed separately for clusters of type A monomers and for
clusters of type B monomers. Otherwise the A and B beads that
belong to the same chain would always belong to a single cluster
containing all the beads of the chain. Similar conclusions have
been also drawn by a similar model~\cite{8a,8b,8c}, where also the
geometry of these cluster phases has been analyzed in detail.
Guided by the theory~\cite{3}, one would rather expect in the long
chain limit that a ground-state-type structure would be a single
lamellar domain, where an interface between all A- and B-type
blocks is formed, similarly to what is known for multiblock
copolymer melts for symmetric choice of parameters. Such a
structure would have much less (unfavorable) A-B contacts compared
to a multidomain structure of A and B clusters, which is
kinetically favored in simulations. Recently, the static
properties of the formed clusters for such systems have been
discussed~\cite{10}, where one is also able to detect in addition
to the phase boundaries the globule-coil crossover and the
proximity of the $\Theta$ temperature through such an analysis. In
this manuscript, we will show further, that properties such as the
bond correlation and asphericity can provide a rather accurate
evidence for detection of full phase separation (single lamellar
structure) in multiblock copolymer chains. Additionally,  we will
argue that these properties can provide information on the length
$N$ and the number of blocks $n$ that this phase behavior is
taking place.

\begin{figure}
\begin{center}
\rotatebox{0}{\resizebox{!}{0.15\columnwidth}{%
    \includegraphics{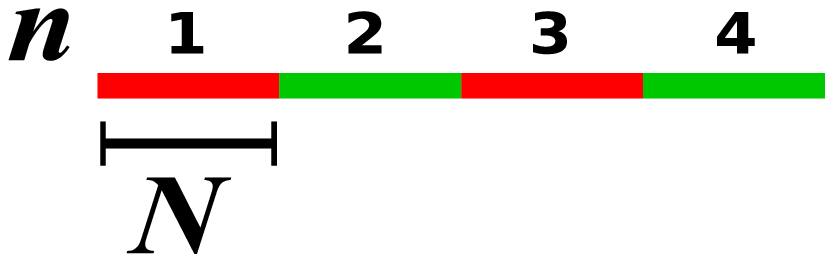}
}}
\end{center}
\caption{\label{fig1}(Color online) Definition of structural
parameters describing our linear multiblock copolymer chains. $n$
(in this case $n=4$) is the number of different blocks A and B
($n_{A}=n_{B}=n/2$) denoted with different color (or grey tone)
and $N$ is the length of each block. All the blocks, irrespective
of whether they are of type A or B, have the same length $N$. Then
the total length of the chain is $nN$.}
\end{figure}

Furthermore, we provide a complete understanding and a broad
discussion of the static properties of such macromolecules under
poor solvent conditions for the blocks and the multiblock chain as
a whole, at least for the most symmetric case that is considered
here, and for chain lengths and thermodynamic conditions
accessible to our simulations. As we have chosen in our study the
same potential parameters as for the standard LJ fluid ~\cite{17},
our results could also serve as a benchmark for similar polymeric
and biological systems. We discuss our results only in the poor
solvent regime, which is the most interesting case due to the
different phase behavior that has been observed when the chain is
in the collapsed state (globule state)~\cite{10,11,12}.

The rest of the paper is organized as follows. In Sec.~\ref{sec:2}
we describe our simulation model and the methods to analyze our
results. In Sec.~\ref{sec:3} we present our results for the whole
multiblock chain and the individual blocks and we discuss
particular details of the studied properties pertinent to
the phase behavior of the multiblock chains. Finally, in
Sec.~\ref{sec:4} we summarize our conclusions in connection with
recent results on the phase behavior of symmetric linear
multiblock copolymers.

\section{MODEL AND METHODS TO ANALYZE THE SIMULATIONS}
\label{sec:2}

As a detailed study of the systems under consideration is
computationally demanding for reasons that will be discussed
below, we have focused only on symmetric linear multiblock
copolymers, where blocks of type A beads and blocks of type B
beads alternate along the chain. Then the length of all blocks,
irrespective of whether being of type A or B, is $N$, the number
of A-type blocks $n_{A}$ is equal to the number of B-type blocks
$n_{B}$, with $n=n_{A} + n_{B}$ an even number denoting the total
number of blocks. A multiblock copolymer chain is described by the
parameters shown schematically in Fig.~\ref{fig1}, where the
different colors correspond to the different types of blocks,
which alternate along the chain.

To model our chains we use the standard bead-spring
model~\cite{18,19,20,21,22,23} to describe a multiblock copolymer
chain, where all beads interact with a truncated and shifted LJ
potential $U_{LJ}(r)$ and nearest neighbors bonded together along
the chain also experience the finitely extensible nonlinear
elastic potential $U_{FENE}(r)$, $r$ being the distance between
the beads. Thus,
\begin{equation}
\label{eq:1}
U_{LJ}^{\alpha \beta}(r)= 4 \varepsilon_{LJ}^{\alpha \beta} [(\sigma_{LJ}^{\alpha \beta}/r)^{12} -
(\sigma_{LJ}^{\alpha \beta}/r)^6] + C, \quad r \leq r_c \quad,
\end{equation}
where $\alpha$, $\beta=A, B$ denote the different type of
monomers, and the constant $C$ is defined such that the potential
is continuous at the cut-off ($r_{c}=2.5$). Henceforth units are
chosen such that $\varepsilon^{\alpha\alpha}=1$,
$\sigma^{\alpha\alpha}=1$, the Boltzmann constant $k_{B}=1$, and
in addition the mass $m^{\alpha\beta}_{LJ}$ of beads is chosen to
be unity. For simplicity, $\sigma_{LJ}^{\alpha \beta}=1$, but
$\varepsilon_{LJ}^{AA}=\varepsilon_{LJ}^{BB}=2\varepsilon_{LJ}^{AB}=1$,
in order to create an unmixing tendency between monomers A and B
belonging to different blocks as done in previous
studies~\cite{10,11,12,24,25,26}, and as is used for a standard
system (LJ fluid)~\cite{17}. Therefore, $\Delta \varepsilon=
\varepsilon_{LJ}^{AB}-1/2(\varepsilon_{LJ}^{AA}+\varepsilon_{LJ}^{BB})$
was kept the same throughout our simulations and $\chi$ ($\chi
\propto \Delta \varepsilon /T$) was mainly varied by tuning the
temperature $T$. The potential of Eq.~(\ref{eq:1}) acts between
any pair of beads, irrespective of whether they are bonded or not.
For bonded beads additionally the potential $U_{FENE}(r)$ acts,
\begin{equation}
\label{eq:2}
U_{\rm FENE} =-\frac{1}{2} k r^2_0 \ln [1-(r/r_0)^2], \quad 0 < r
\leq r_0.
\end{equation}
where $U_{FENE}(r\geq r_{0})= \infty$, and hence $r_0$ is the
maximal distance between bonded beads. We use the standard choice
$r_0=1.5$ and $k=30$. We recall that for linear chains the
$\Theta$ temperature for the present model has been roughly
estimated as $T_{\Theta} \approx 3.0$ (a more accurate
value~\cite{27} $T_{\Theta} \approx 3.18$ could only be
established for chain length $nN>200$, $n=1$). Thus, we have
considered in our study the range of temperatures $1.5 \leq T \leq
3.0$. Moreover, in previous discussion of this system~\cite{10} we
showed that the chain enters in the globule regime at temperature
$T \approx 2.4$. Therefore, phase separation between A and B
monomers takes place below this temperature. For homopolymer
chains this boundary is sharper compared to the case of multiblock
copolymers where also appears smoother as the number of blocks $n$
increases for constant chain length $nN$ as the A-B junctions
between blocks increase. Of course, the temperature that we enter
the globule regime (upon lowering the temperature) is the same for
homopolymer and copolymer chains, because in our model the
location of this boundary depends only on the solvent (controlled
by temperature in our simulations), which is the same for a
homopolymer chain, as well as for type A and type B beads of a
copolymer chain, but in the case of copolymer chains it only appears
smoother.

Further on our simulation method, the positions $\vec{r}_i(t)$ of
the effective monomers with label $i$ evolve in time $t$ according
to Newton's equation of motion, amended by the Langevin thermostat
\begin{equation}
\label{eq:3} m \frac {d^2\vec{r}_i}{dt^2} = - \nabla U_i-\gamma
\frac {d\vec{r}_i}{dt} + \vec{\Gamma}_i(t),
\end{equation}
where $U_i({\vec{r}_{j}(t)})$ is the total potential acting on the
$i$'th bead due to its interactions with the other beads at sites
$\vec{r}_{j}(t)$, $\gamma$ is the friction coefficient, and
$\vec{\Gamma}_i(t)$ the associated random force. The latter is
related to $\gamma$ by the standard fluctuation-dissipation
relation
\begin{equation}\label{eq:4}
\langle \vec{\Gamma}_i(t)\cdot \vec{\Gamma}_j(t')\rangle = 6 k_BT
\gamma \delta _{ij}\delta (t-t')\;.
\end{equation}
Following previous work~\cite{18,19,20,21,22,23,24,25,26,27}, we
choose $\gamma=0.5$ , the MD time unit $\tau_{LJ} = (m \sigma
^2_{LJ}/\varepsilon^{\alpha\alpha}_{LJ})^{1/2}=1$ being also unity
for our choice of units. Equation~(\ref{eq:3}) is integrated using
the leap-frog algorithm~\cite{28}, with a time step $\Delta t =
0.006 \tau_{LJ}$ and utilizing the GROMACS package~\cite{29}.

Simulation of polymer chains under poor solvent conditions is
notoriously difficult as relaxation times become exceedingly high.
Therefore, we describe here briefly our simulation procedure.
After equilibration at a ``high'' temperature (in the coil
regime), we collect a number of independent samples (typically
$500$), which we use as initial configurations for slow cooling
runs, as is done in previous studies~\cite{24,25,26}. For longer
chains, temperatures higher than $T=3.0$ were used in order to
facilitate the procedure of obtaining initial independent samples.
We note here that the solvent is taken into account in our model
only implicitly by tuning the temperature~\cite{10,11,12}. Then,
decrease of the temperature corresponds to higher incompatibility
of the implicit solvent with the monomers. For each cooling
history, we lower the temperature in steps of $\Delta T=0.1$ and
run the simulations for $2 \times 10^6$ MD steps using the final
configuration at each (higher) temperature as starting
configuration for the next (lower) temperature. At low enough
temperatures~\cite{11} (typically $T<2.0$, at a rather high
distance from the $\Theta$ temperature), where dense ``clusters''
of a few neighboring blocks are formed, it is not possible to run
simple MD simulations long enough to sample the phase space
adequately. Therefore, using this procedure of independent cooling
histories is indispensable for obtaining meaningful statistics.
This large statistical effort prevents the study of exceedingly
long chains for this simulation model.

At temperatures close to $\Theta$, coil-like structures are formed
and the differences between multiblock copolymer chains with
different $n$, $N$ are rather small, expecting that at $T
\rightarrow \infty$ these differences will completely disappear.
At temperatures below $T \approx 2.4$ the chains collapse and
blocks of the same type of monomers form clusters with other
monomers belonging to different blocks, due to the unfavorable
interactions between A and B beads~\cite{10,11,12}. The static
properties of these clusters have been described in detail in
previous communication~\cite{10}. Here, we focus on the static
properties of the whole chain (of length $nN$) as well as the
static properties of individual blocks of $N$ effective monomers.
Due to the symmetric choice of our parameters, the averages of the
static properties for A and B blocks should not differ. Therefore,
the calculation for averages of the properties of the blocks
should be taken over all blocks $n$, irrespective of their type,
and their properties will be denoted with the subscript ``b'', for
example the average value of the gyration radius of individual
blocks would read $R_{g,b}$ while those for the whole chains
without this index, namely as $R_{g}$. We discuss below the
properties that might need some more details.

\begin{figure}[]
\begin{center}
\subfloat[][]{
\rotatebox{270}{\resizebox{!}{1.00\columnwidth}{%
   \includegraphics{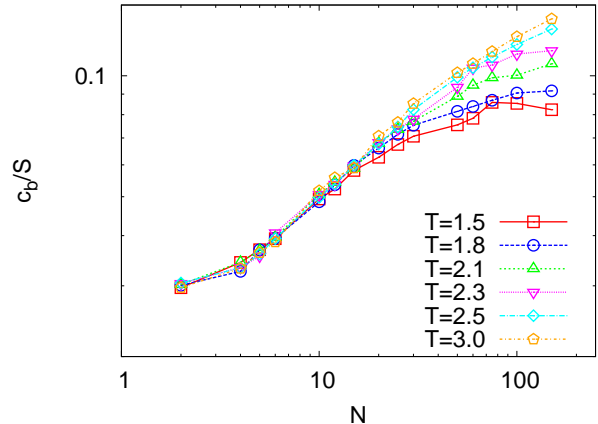}
}}}\\
\subfloat[][]{
\rotatebox{270}{\resizebox{!}{1.00\columnwidth}{%
   \includegraphics{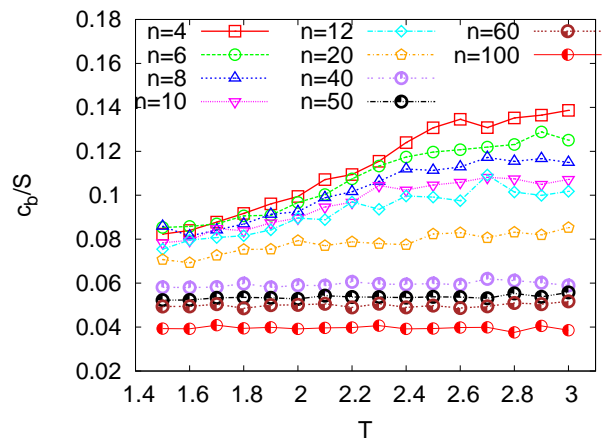}
}}}
\end{center}
\caption{\label{fig2}(Color online) Acylindricity of the blocks
for chains of total lentgh $nN=600$. The dependence with the block
length $N$ is shown for different temperatures (a) and the
dependence with the temperature $T$ for different number of blocks
$n$ (b).}
\end{figure}

We follow the definition of standard textbooks~\cite{30,31,32} to
discuss aspects related to the persistence length of the polymer
chains, where the persistence length is defined from the decay of
bond orientational correlations along the chain. However, one
should be careful when defines the persistence length with this
correlation function as it has been discussed
elsewhere~\cite{33,34}. Nevertheless, we show in this work that
this property is extremely accurate in indicating the appearance
of full phase separation of the blocks in symmetric multiblock
copolymers. We give here a brief definition of this property.
Defining the bond vectors $\vec{\alpha}_{i}$ in terms of the
monomer positions $\vec{r}_{i}$ as
$\vec{\alpha}_{i}=\vec{r}_{i+1}-\vec{r}_{i}$, $i=1,...,nN-1$, the
bond orientational correlation is defined as
\begin{equation}
\label{eq:5}
\langle cos \theta(s) \rangle= l_{b}^{-2}  \frac{1}{N_{b}-1-s}  \sum_{i=1}^{N_{b}-1-s} \langle \vec{\alpha}_{i} \cdot \vec{\alpha}_{i+s} \rangle.
\end{equation}
Note $\vec{\alpha}_{i}^2=l_{b}^{2}$, with $l_{b}$ being the
average bond length. For this model $l_{b} \approx 1.00$ and hence
$\langle cos \theta(0) \rangle \approx 1$, of course. Considering
the limit $nN \rightarrow \infty$ and assuming Gaussian chain
statistics, one obtains an exponential decay, since then $\langle
cos\theta(s) \rangle= \langle cos\theta(1) \rangle^{s}=
\exp[\ln\langle cos\theta(1) \rangle]$, and thus
\begin{equation}
\label{eq:6} \langle cos\theta(s) \rangle= \exp[-s/l_{p}],
l_{p}^{-1}=-\ln\langle cos\theta(1) \rangle.
\end{equation}
For chains at the $\Theta$ point~\cite{35} or in melts~\cite{36}
\begin{equation}
\label{eq:7} \langle cos\theta(s) \rangle \propto s^{-3/2},
s\rightarrow \infty.
\end{equation}
Below $\Theta$ temperature this correlation function falls-off
rapidly at small $s$ depending strongly on temperature $T$.

In our study we also consider properties related to the shape of
multiblock copolymers. We follow the description of Theodorou and
Suter~\cite{37} to define the apshericity and acylindricity of a
mutliblock copolymer chain. First, we define the gyration tensor
as
\begin{equation}
\label{eq:8}
\mathbf{S} = \frac{1}{N_{\xi}} \sum_{i=1}^{N_{\xi}} \mathbf{s}_{i} \mathbf{s}_{i}^{T}= \overline{\mathbf{s} \mathbf{s}^{T}}=
\left[ {\begin{array}{ccc}
  \overline{x^2} & \overline{xy} & \overline{xz} \\
  \overline{xy} & \overline{y^2} & \overline{yz} \\
  \overline{xz} & \overline{yz} & \overline{z^2} \\
 \end{array} } \right],
\end{equation}
where $\mathbf{s}_i=col(x_i,y_i,z_i)$ is the position vector of
each bead, which is considered with respect to the center of mass
of the beads $\sum_{i=1}^{N_{\xi}} \mathbf{s}^i=0$, and the
overbars denote an average over all beads $N_{\xi}$. When the
gyration tensor of the whole chain is considered, then
$N_{\xi}=nN$. For the blocks $N_{\xi}=N$, the gyration tensor and
the properties are calculated for each block separately and then
an average over all blocks irrespective of their type is taken. As
mentioned earlier, this is perfectly valid due to the symmetry of
our model. The gyration tensor is symmetric with real eigenvalues
and a cartesian system that this tensor is diagonal can always be
found, i.e.,
\begin{equation}
\label{eq:9} \mathbf{S}=diag(\overline{X^2}, \overline{Y^2},
\overline{Z^2}),
\end{equation}
where the axes are also chosen in such way that the diagonal
elements (eigenvalues of $\mathbf{S}$) $\overline{X^2}$,
$\overline{Y^2}$,  and $\overline{Z^2}$ are in descending order
($\overline{X^2} \ge \overline{Y^2} \ge \overline{Z^2}$). These
eigenvalues are called the principal moments of the gyration
tensor. From the values of the principal moments, one defines
quantities such as the asphericity $b$,
\begin{equation}
\label{eq:10}
b= \overline{X^2} -1/2( \overline{Y^2} + \overline{Z^2} ).
\end{equation}
When the particle distribution is spherically symmetric or has a
tetrahedral or higher symmetry, then $b=0$. The acylindricity $c$
\begin{equation}
\label{eq:11} c=\overline{Y^2} - \overline{Z^2}
\end{equation}
is zero when the particle distribution is in sync with a
cylindrical symmetry. Therefore, the acylindricity and asphericity
are the relevant quantities that would describe geometries that
could be relevant in the case of a homopolymer or multiblock
chain. These quantities are taken with respect to $S$, to the sum
of the eigenvalues, i.e. the square gyration radius of the chain,
which we also have calculated independently on our original
cartesian coordinates in order to check our results.

Assuming an ellipsoidal shape and based on the calculation of the
above eigenvalues of the gyration tensor, one can define an
effective volume expressed by the following relation~\cite{38}
\begin{equation}
\label{eq:12} V^{eff}=4\pi \sqrt[]{3} \prod_{i=1}^{3}
\sqrt[]{\lambda_{i}},
\end{equation}
where $\lambda_{1}= \overline{X^2}$, $\lambda_{2}=\overline{Y^2}$,
and $\lambda_{3}=\overline{Y^2}$, are the eigenvalues of the
radius of gyration tensor. Then, the effective radius of a sphere
with the same volume as this ellipsoid ($V^{eff}$) is given by the
geometrical mean of individual radii $R_{g}^{eff}= \sqrt[]{3}
\prod_{i=1}^{3} \sqrt[6]{\lambda_{i}}$. This can be compared with
the volume of an effective sphere defined by the gyration radius
$R_{g}=\sqrt[]{\sum^{i=1}_{3} \lambda_{i}}$. Such an analysis is
not in the scope of the present study.

\begin{figure}[]
\begin{center}
\rotatebox{270}{\resizebox{!}{1.00\columnwidth}{%
   \includegraphics{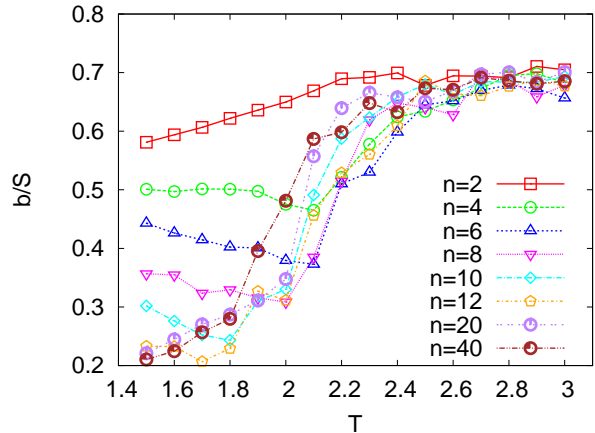}
}}
\end{center}
\caption{\label{fig3}(Color online) The dependence of asphericity
with the temperature $T$ for different combinations of $n$, $N$ for
chains of total length $nN=600$ is shown.}
\end{figure}

Another property that provides information on the microscopic
properties of multiblock chains is the number of contacts between
beads of the same or different type. Smaller number of contacts
A-B suggests a tendency towards phase separation. The number of
contacts can be mathematically described by the following
formula~\cite{10,26}
\begin{equation}
\label{eq:13}n_{\alpha \beta} = 4 \pi\int\limits_{0}^{r_{n}}
{g}_{\alpha \beta} (\Delta r) (\Delta r)^{2}d(\Delta r),
\end{equation}
where $\Delta r$ is the absolute value of the distance between two
sites of monomers $\vec{r}_{i},\vec{r}_{j}$ in the multiblock
copolymer chain, and $g_{\alpha \beta}$ the corresponding radial
distribution function. Equation~(\ref{eq:13}) means that a pair of
monomers [$\alpha$, $\beta$, $(\alpha,\beta=A,B)$] is defined to
have a pairwise contact if their distance is less than $r_{n}$.
For the distance $r_{n}$ we have used the standard
Stillinger~\cite{stil} neighborhood criterion for monomers. We
followed the standard choice $r_{n}=1.5 \sigma^{\alpha
\beta}_{LJ}$ and checked that qualitatively very similar results
were obtained if one chooses $r_n$ a bit smaller than this choice
(larger values of $r_n$ are physically hardly significant, since
then the particles are too weakly bound, due to the rapid fall-off
of the LJ potential). Therefore, a pair of monomers being an
absolute distance less than $r_{n}=1.5$ apart define a
``contact''. Then, the numbers presented in this manuscript will
denote the average number of neighbors per monomer.

\begin{figure}[]
\begin{center}
\subfloat[][]{
\rotatebox{270}{\resizebox{!}{1.00\columnwidth}{%
   \includegraphics{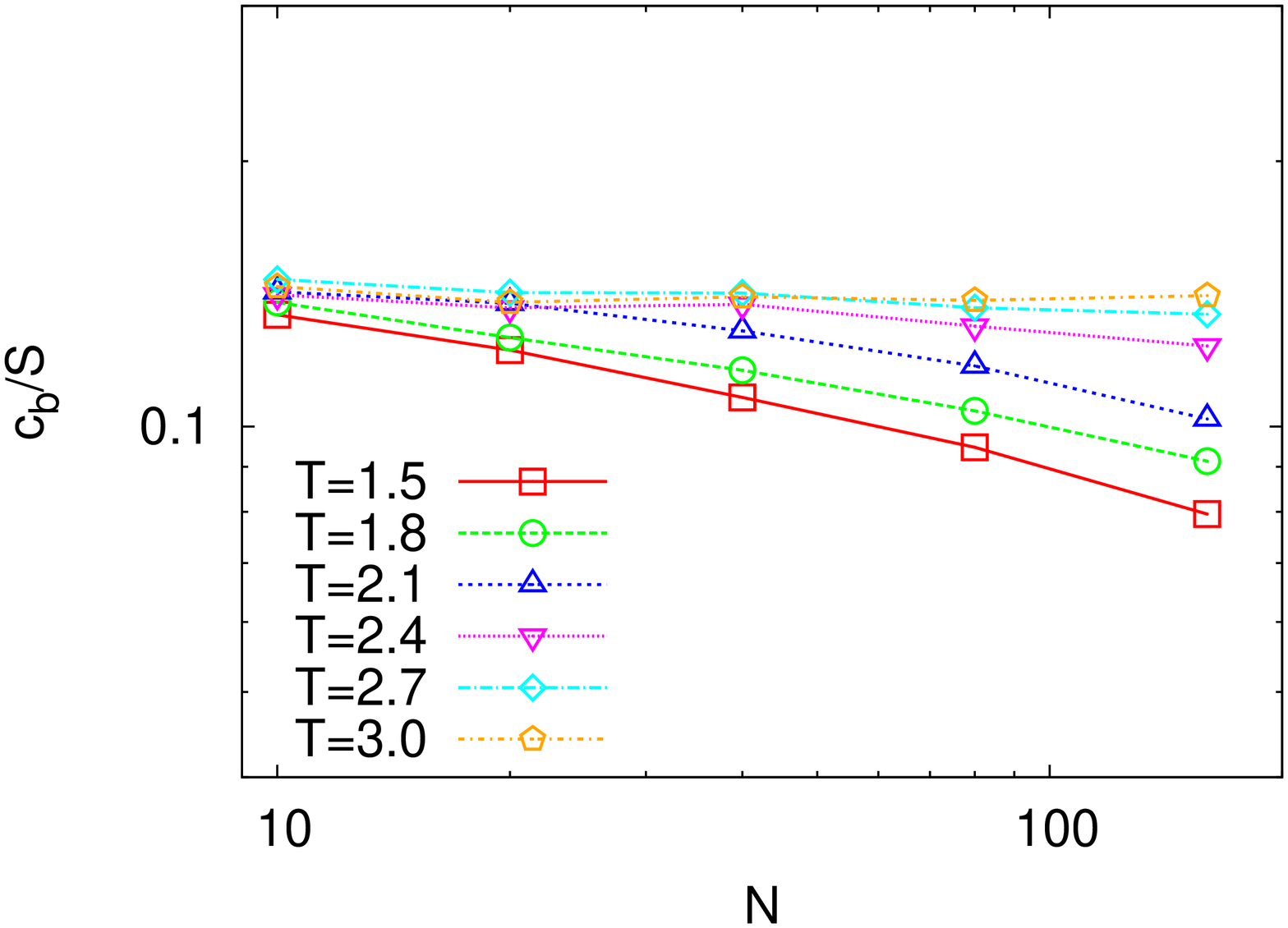}
}}}\\
\subfloat[][]{
\rotatebox{270}{\resizebox{!}{1.00\columnwidth}{%
   \includegraphics{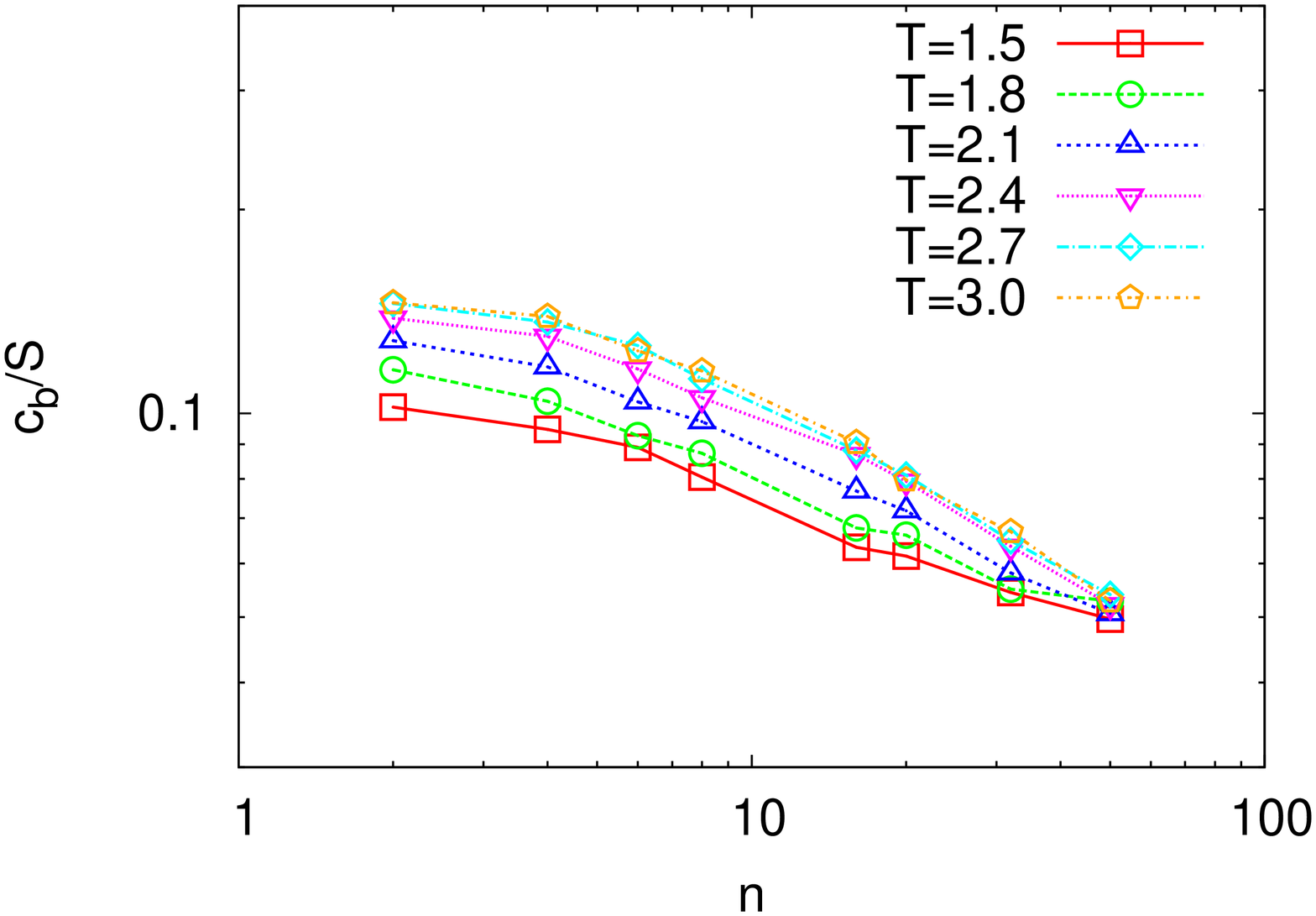}
}}}
\end{center}
\caption{\label{fig4}(Color online) (a) Acylindricity of the the
blocks of a chain with $n=8$ blocks as a function of the block
length $N$ for different temperatures, as indicated. The
dependence of acylindricity on the number of blocks $n$ is shown
correspondingly in part (b) for
$N=80$.}
\end{figure}

\begin{figure}[]
\begin{center}
\subfloat[][]{
\rotatebox{270}{\resizebox{!}{1.00\columnwidth}{%
   \includegraphics{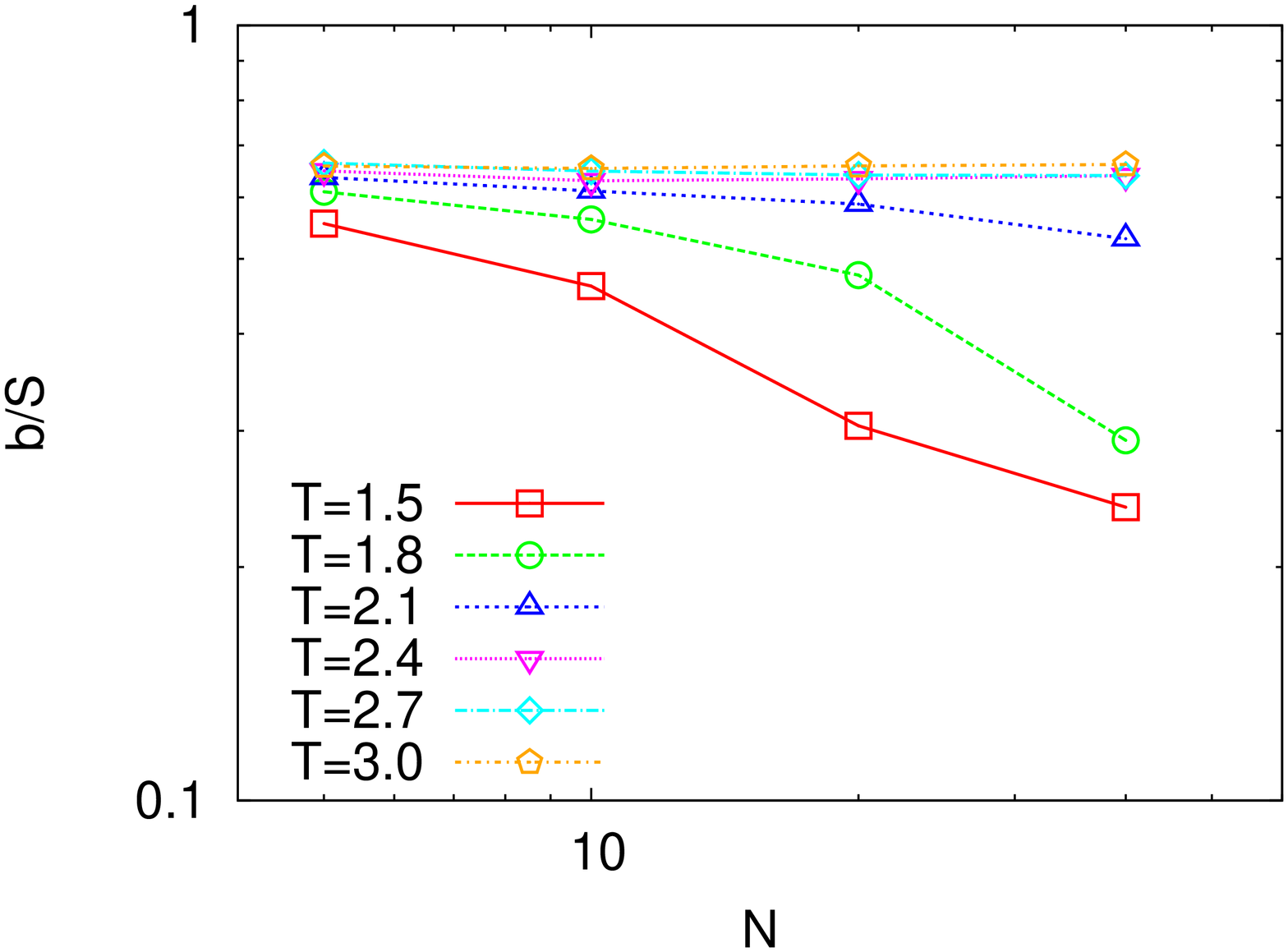}
}}}\\
\subfloat[][]{
\rotatebox{270}{\resizebox{!}{1.00\columnwidth}{%
   \includegraphics{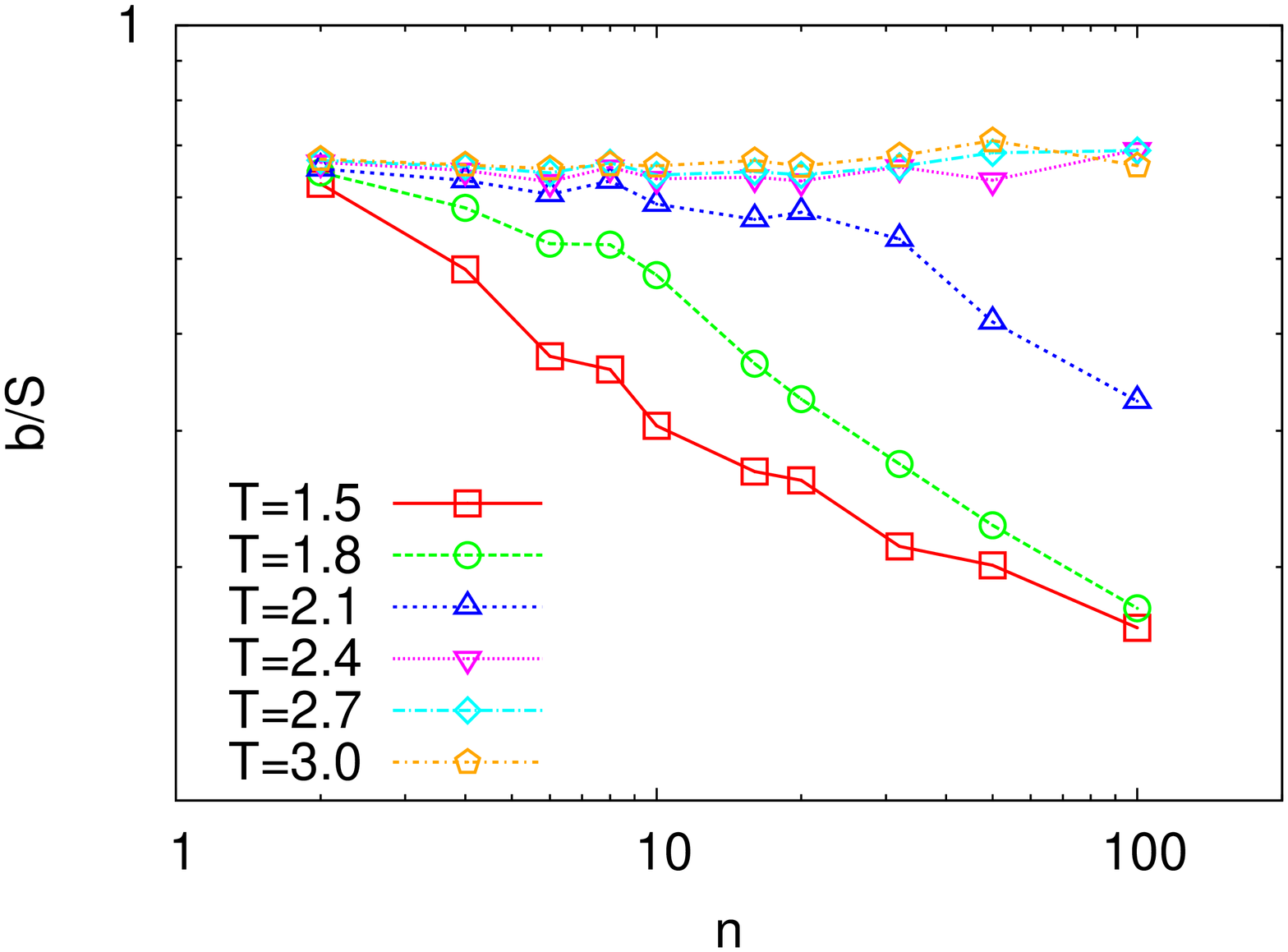}
}}}
\end{center}
\caption{\label{fig5}(Color online)
Asphericity of chains composed of $n=10$ blocks as
a function of the block length $N$ (a), and asphericity
of chains with block length $N=20$ as a function of
the number of blocks $n$ (b).
}
\end{figure}

\begin{figure}[]
\begin{center}
\subfloat[][]{
\rotatebox{270}{\resizebox{!}{1.00\columnwidth}{%
   \includegraphics{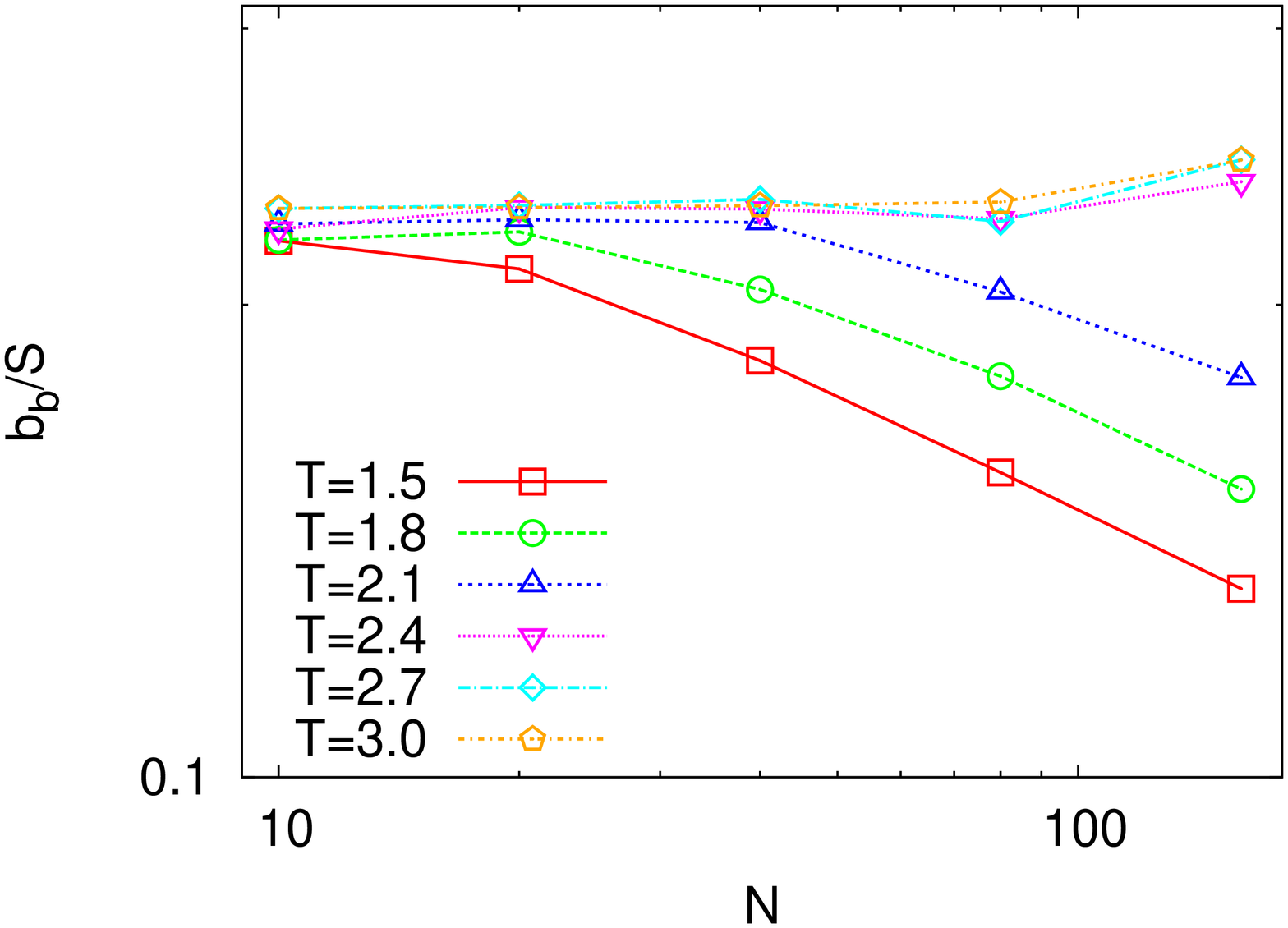}
}}}\\
\subfloat[][]{
\rotatebox{270}{\resizebox{!}{1.00\columnwidth}{%
   \includegraphics{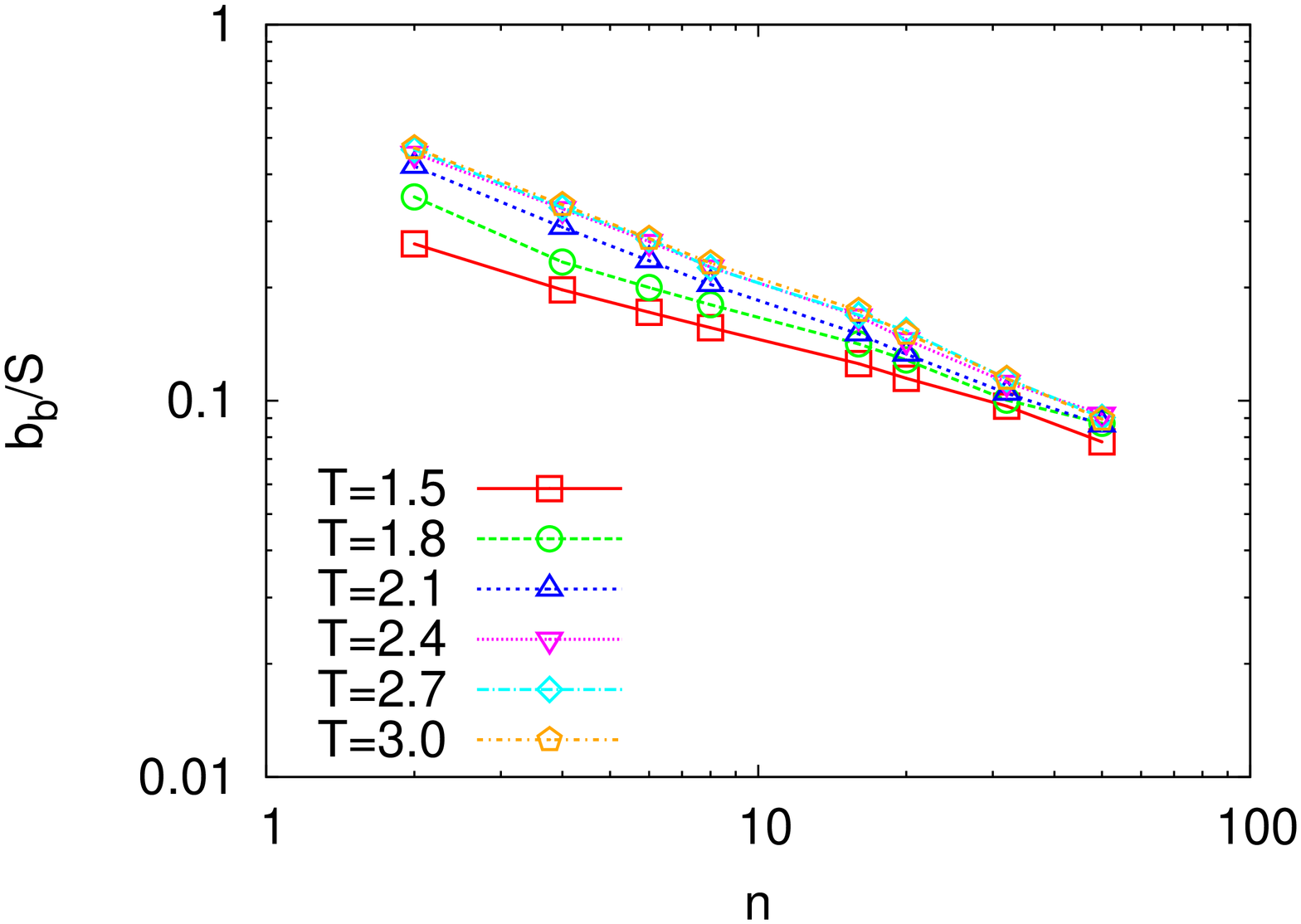}
}}}
\end{center}
\caption{\label{fig6}(Color online)
Asphericity of the blocks of multiblock chains with $n=8$ as
a function of the block length $N$ (a), and of
chains with block length $N=80$ (b)
as a function of the number of blocks $n$.
}
\end{figure}

\begin{figure}[]
\begin{center}
\subfloat[][]{
\rotatebox{270}{\resizebox{!}{1.00\columnwidth}{%
   \includegraphics{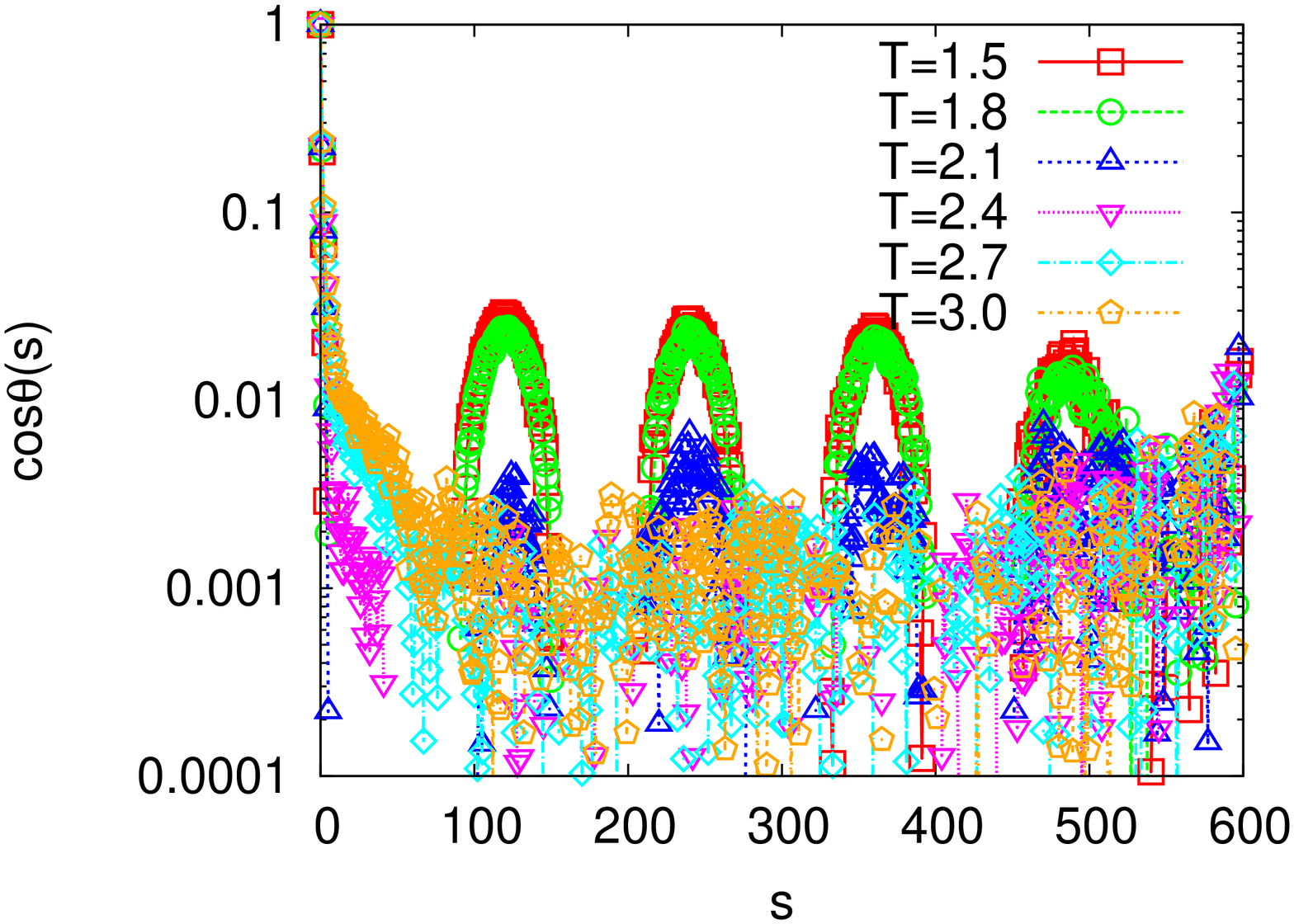}
}}}\\
\subfloat[][]{
\rotatebox{270}{\resizebox{!}{1.00\columnwidth}{%
   \includegraphics{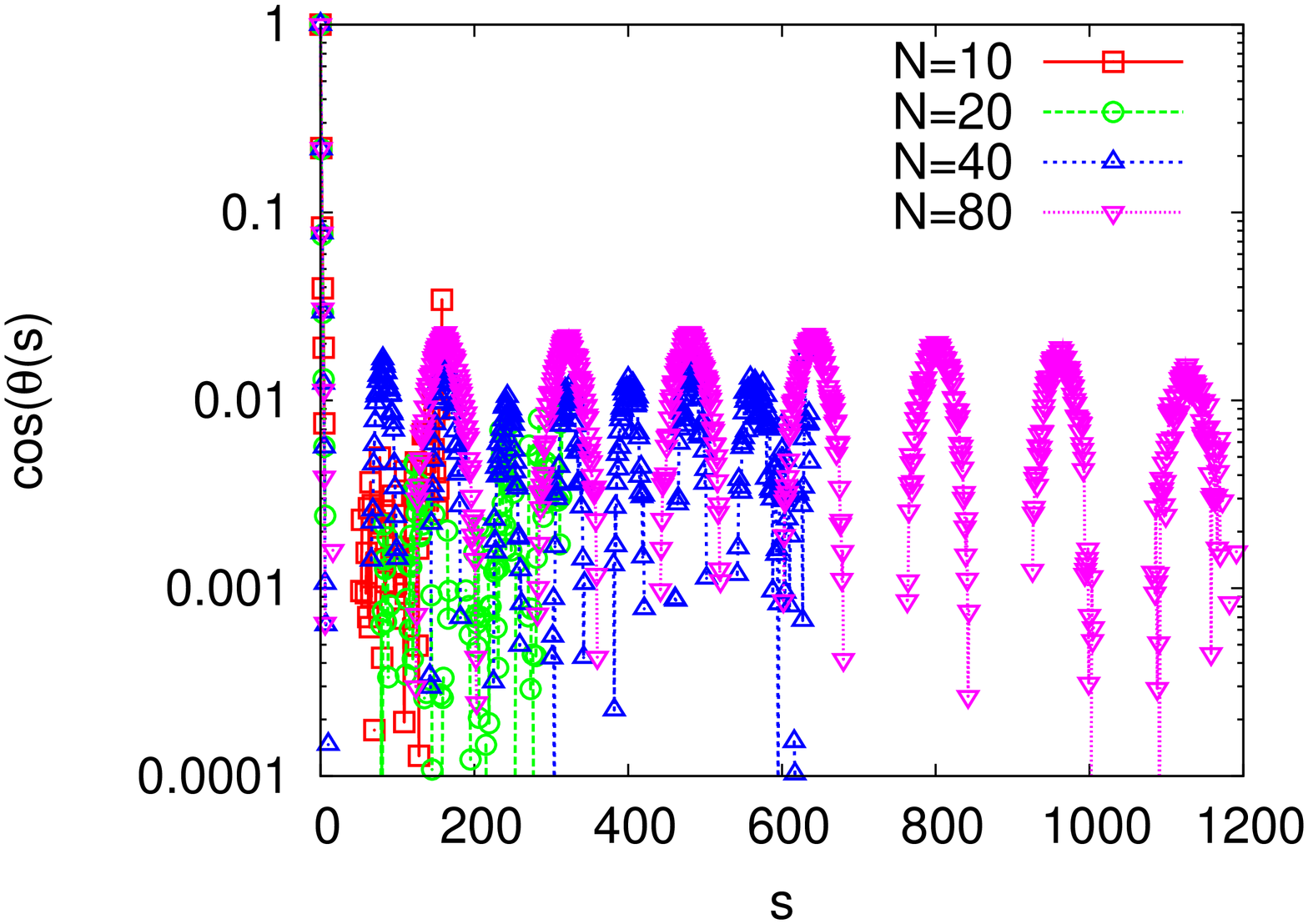}
}}}
\end{center}
\caption{\label{fig7}(Color online) Bond orientational correlation
for the case $n=10$, $N=60$ (a) corresponding to a full phase
separated system at temperature below $T=2.4$ (here $T=1.5$), i.e.
all blocks of type A join in one cluster while the blocks of type
B form another cluster. At
temperatures $T>2.4$ the chain obtains coil-like formations.
The dependence on the
block length $N$ for chains with $n=16$ is shown in part (b). For
block length $N>20$ the phase separation between blocks A and B
takes place. These data refer to the temperature $T=1.8$. }
\end{figure}

\section{RESULTS}
\label{sec:3}

Firstly, we discuss properties related to the acylindricity and
asphericity of the chains. The acylindricity defines the deviation
of the shape of the chain from a cylindrical geometry;
correspondingly the asphericity expresses the deviation from a
spherical shape or tetrahedral or higher symmetry with respect to
the total dimensions of the chains. Then Fig.~\ref{fig2}a presents
the dependence of the acylindricity of individual blocks on
average for multiblock copolymer chains of the same total length
$nN=600$ on the block length $N$. Thus, the number of blocks for
each chain is $n=600/N$. This choice for the total length of the
multiblock chain allows for the observation of all interesting
regimes related to the phase behavior of such macromolecules
\cite{10,11,12}. In Fig.~\ref{fig2}a one can observe very clearly
a regime for intermediate values of $2<N<20$ where the curves show
small deviations from one another. The cases of $N=2$ and $N=1$
are particular. For $N=1$ (not shown) the acylindricity drops to
zero. At lower temperatures larger deviations for higher values of
$N$ are seen. Nevertheless from such a plot we are not able to
find any hint that relates to the well-known behavior of these
systems~\cite{10,11,12}. Looking at the dependence of the above
systems with different $N$ on the temperature $T$
(Fig.~\ref{fig2}b) it is clearer that small blocks show very small
dependence on the temperature due to the very small size of the
blocks ($N=600/n$), as it also happens in the plot of
Fig.~\ref{fig2}a. The graph of Fig.~\ref{fig2}b indicates that for
small values of $N$, i.e. $n>20$, the blocks rather do not show
any dependence on the temperature, while for $n<20$ ($N>30$),
stronger dependencies on the the temperature $T$ are seen. This
suggests that stronger deviations from a cylidrical symmetry
persist for multiblock chains of higher $N$, as $nN$ remains
constant. This behavior is also related to the behavior of the
size of the cluster formations in these systems, which has been
discussed elsewhere~\cite{10}. From the results of Fig.~\ref{fig2}
we do not see any direct correlation to the phase behavior of the
multiblock chains by examining the shape of the individual blocks.

On the other hand, the asphericity of the whole multiblock chain
of total length $nN=600$ shows interesting behavior when a plot
with the temperature is attempted. We remind the reader that the
total chain length $nN=600$ is high enough to consider all the
different cases of phase behavior of symmetric multiblock
copolymers, which have already been discussed in
detail~\cite{11,12}. Figure~\ref{fig3} then illustrates two
different regimes. At rather high temperatures ($T>2.4$) the
different multibock chains (of the same total length $nN$) exhibit
small differences as we are close to $\Theta$ conditions. Such
small differences are related to the number of proxima A-B
interactions at the junctions that connect the consecutive blocks
A-B blocks. The values on the $y$-axis of this plot indicate
strong deviation of the chain's shape from a spherical one, due to
the excluded volume interactions in combinations with the
incompatibility between monomers of type A and B at the junctions
of A and B blocks. At high values of $N$ (small $n$ except for the
case $n=2$ which will be separately discussed below) we observe
strong differences at low temperatures. The case $n=4$ reaches at
temperature close to $T=2.2$ a plateau-like regime. We can even
observe a slight increase as temperature decreases and the
incompatibility of A and B monomers increases. As $n$ increases
($N$ decreases) this plateau-like dependence appears
correspondingly at lower temperature $T$. Knowing already the
phase behavior of these systems~\cite{10,11,12}, these
temperatures correspond to those that the multiblock copolymer
chain separates in two clusters with a single A-B interface
forming between them, where each cluster contains only beads of
one type with a permanent interface A-B separating the two
clusters. The transition of multiple A-B interfaces to a single
A-B interface becomes also noticeable in the behavior of
asphericity, for high $n$, i.e. $n>20$, where this plateau now has
disappeared. Therefore, the phase behavior of the multiblock
chains is nicely reflected to a quantity that describes the
overall formation of the chain, without taking into account the
measurement of any microscopic quantities which relate to the
incompatibility between A and B beads.

For the case $n=2$ this transition to the phase separated clusters
can not be concluded from the results of Fig.~\ref{fig3}, because
in this case the chain always has only these two clusters of A and
B beads either in a coil like formation at higher temperatures or
a globule-like formation at lower temperatures. In this case of
$n=2$ we rather see a small decrease in the quantity $b/S$, as the
chain obtains a globular formation. Also, the values of $b/S$ for
$n=2$ are higher than the other cases indicating the stronger
presence of A-B interface. In this case a smaller number of
unfavorable contacts A-B exist (Fig.4b of Ref.~\cite{10}). For all
$n$ shown in Fig.~\ref{fig3} a more spherical formation is
obtained at lower temperatures. In particular this spherical
symmetry is strongly present for the multiblock chains of small
$N$, where also we do not have the appearance of this large A-B
interface and therefore the curves vary smoothly with the
temperature $T$. The above description is valid for all cases
where full phase separation occurs without restrictions on the
length $N$ or number of blocks $n$ and reflects totally the full
phase separation behavior of multiblock copolymers.

\begin{figure}[]
\begin{center}
\subfloat[][]{
\rotatebox{270}{\resizebox{!}{1.00\columnwidth}{%
   \includegraphics{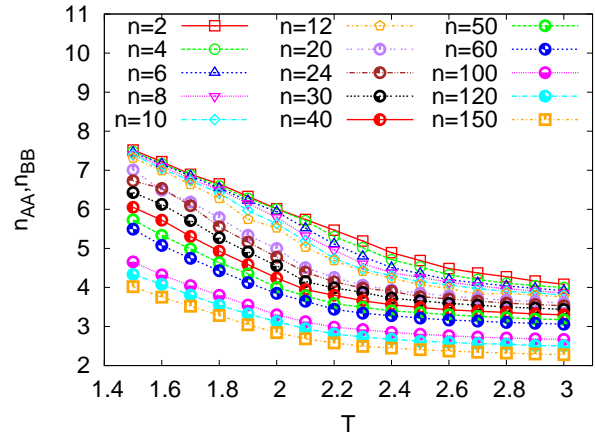}
}}}\\
\subfloat[][]{
\rotatebox{270}{\resizebox{!}{1.00\columnwidth}{%
   \includegraphics{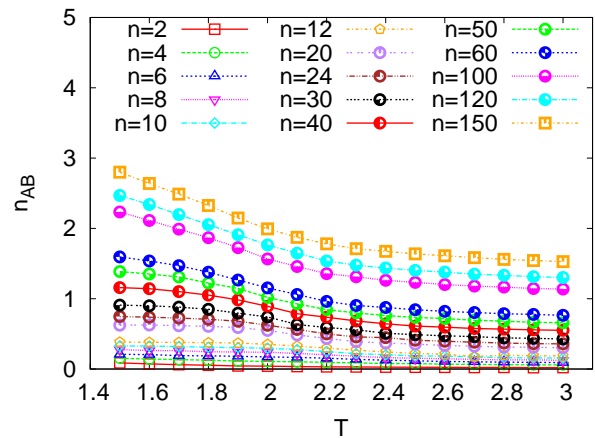}
}}}\\
\subfloat[][]{
\rotatebox{270}{\resizebox{!}{1.00\columnwidth}{%
   \includegraphics{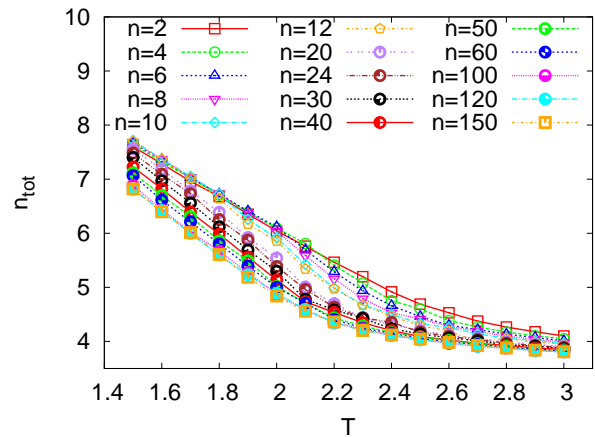}
}}}
\end{center}
\caption{\label{fig8}(Color online) The dependence of the number
of contacts on temperature for different $n$ for chains of total
length $nN=600$. Contacts AA (a), AB (b), and the total number of
contacts irrespective of whether they are of case A or B (c) are
shown. The fraction of A and B monomers is the same for cases, as
the total length of the multiblock chain is kept constant and the
most symmetrical case is considered, i.e. an even number of blocks
of the same length alternate along the chain. Block lengths
fulfilling this condition for $nN=600$ are considered. }
\end{figure}

Now we turn our focus on acylindricity and asphericity for the
blocks and the whole multiblock chains on their dependence with
the block length $N$ and the number of blocks $n$, so without the
fact that results correspond to chains of the same total length
$nN$. Figure~\ref{fig4}a shows such results for the average
acylindricity of individual blocks and its dependence on the block
length $N$ for a multiblock chain containing $n=8$ blocks. It has
been shown that for this number of blocks phase separation in two
clusters A and B is taking place for high enough block lengths
($N>20$) following our simulation protocol. However, the plot of
this quantity for individual blocks does not reflect the different
phase separation behavior for such systems. At temperatures close
to the $\Theta$, i.e. $T=3.0$, an almost straight line is seen
suggesting that chain blocks appear with the same overall
formation irrespective of their block length $N$. At lower
temperatures down to $T=1.5$ deviations are seen from the above
behavior and the chains with higher $N$ show a higher dependence
on the temperature $T$. Larger blocks tend to obtain a more
cylindrical-like shape as the temperature decreases at $T=1.5$, something
that is seen for all cases.

However, it becomes clear that even for small $N$ the increase of
the number of blocks $n$ favors a more cylindrical shape of
individual blocks on average. In the case that the chain length
$N$ is high enough that full phase separation can take place
($N=80$, Fig.~\ref{fig4}b), i.e., one cluster contains the beads of
all blocks of type A and another cluster beads of all blocks of
type B, variations due to temperature can become now more apparent.
For $N=80$ we observe the same tendency that
is seen for $N=5$, but the chains are long enough to observe the
differences attributed to different temperatures. This can also be
concluded by examining the $n$ dependence of multiblock chains
with various block lengths $N$. The graphs of Fig.~\ref{fig4} show
that the shape of individual blocks is affected by the number and
block of lengths, but phenomena of phase separation can not be
identified on an individual block-based analysis as a function of
block length $N$ or number of blocks $n$.

\begin{figure}[]
\begin{center}
\subfloat[][]{
\rotatebox{270}{\resizebox{!}{1.00\columnwidth}{%
   \includegraphics{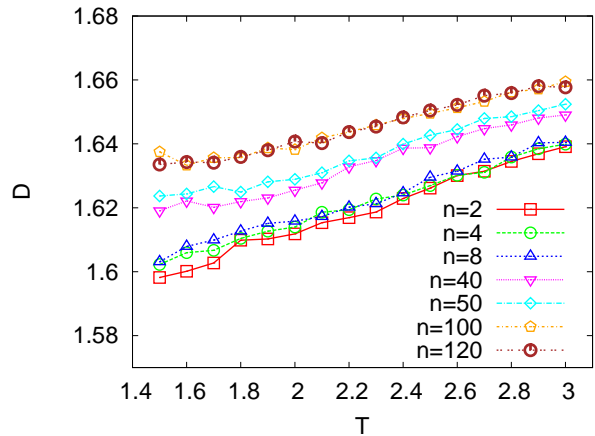}
}}}\\
\subfloat[][]{
\rotatebox{270}{\resizebox{!}{1.00\columnwidth}{%
   \includegraphics{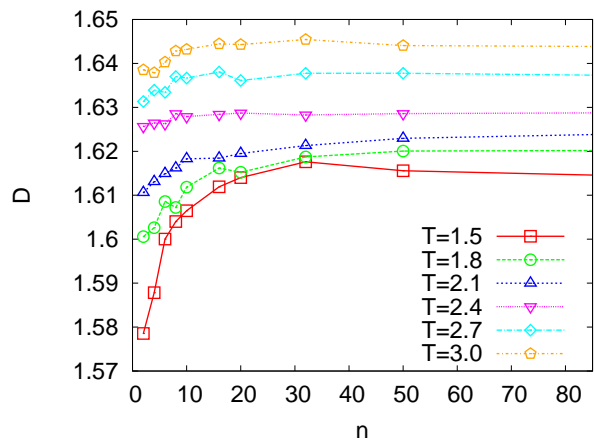}
}}}\\
\subfloat[][]{
\rotatebox{270}{\resizebox{!}{1.00\columnwidth}{%
   \includegraphics{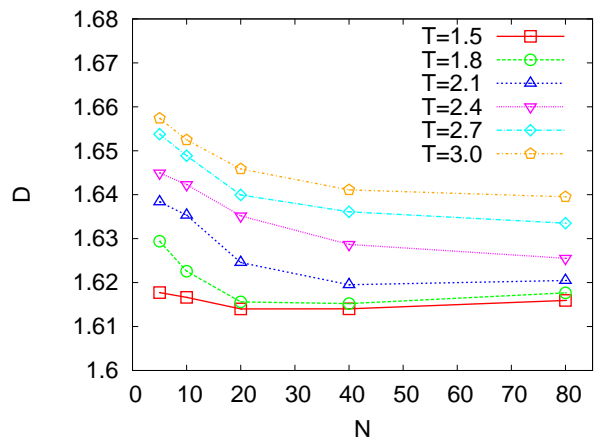}
}}}
\end{center}
\caption{\label{fig9}(Color online) The dihedral angle as a
function of temperature for chains of total length $nN=600$ (a).
The dependence of the same property for chains with $N=40$ with
the number of blocks $n$ (b), and for chains with $n=20$ with the
block length $N$ (c). $D$ is measured in radians. }
\end{figure}

\begin{figure}[]
\begin{center}
\subfloat[][]{
\rotatebox{270}{\resizebox{!}{1.00\columnwidth}{%
   \includegraphics{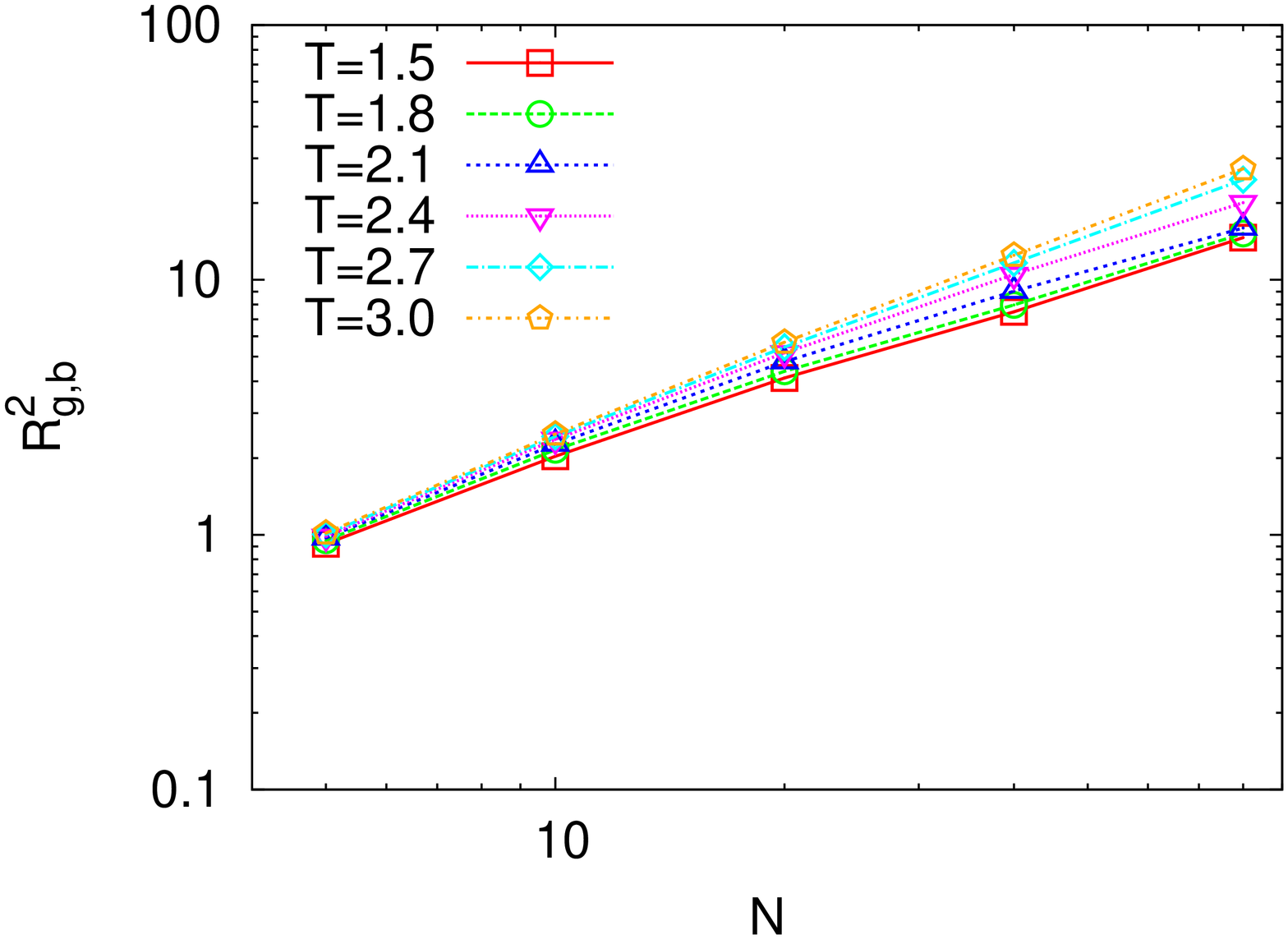}
}}}\\
\subfloat[][]{
\rotatebox{270}{\resizebox{!}{1.00\columnwidth}{%
   \includegraphics{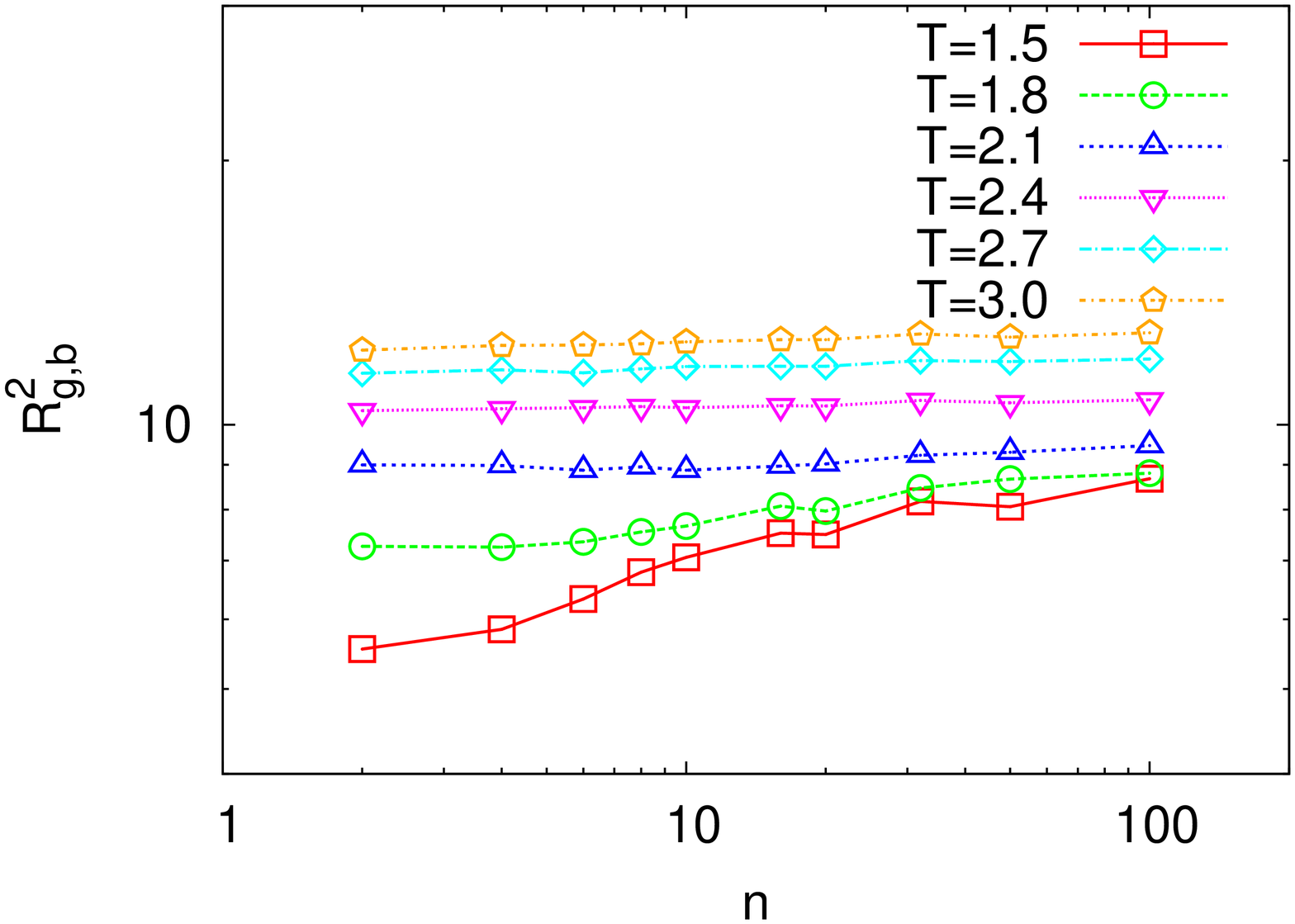}
}}}
\end{center}
\caption{\label{fig10}(Color online) Two representative cases for
the mean square gyration radius of the blocks for chains with
$n=20$ (a), and for chains with $N=40$ (b) for different
temperatures as indicated.}
\end{figure}

\begin{figure*}[]
\begin{center}
\subfloat[][]{
\rotatebox{270}{\resizebox{!}{1.00\columnwidth}{%
   \includegraphics{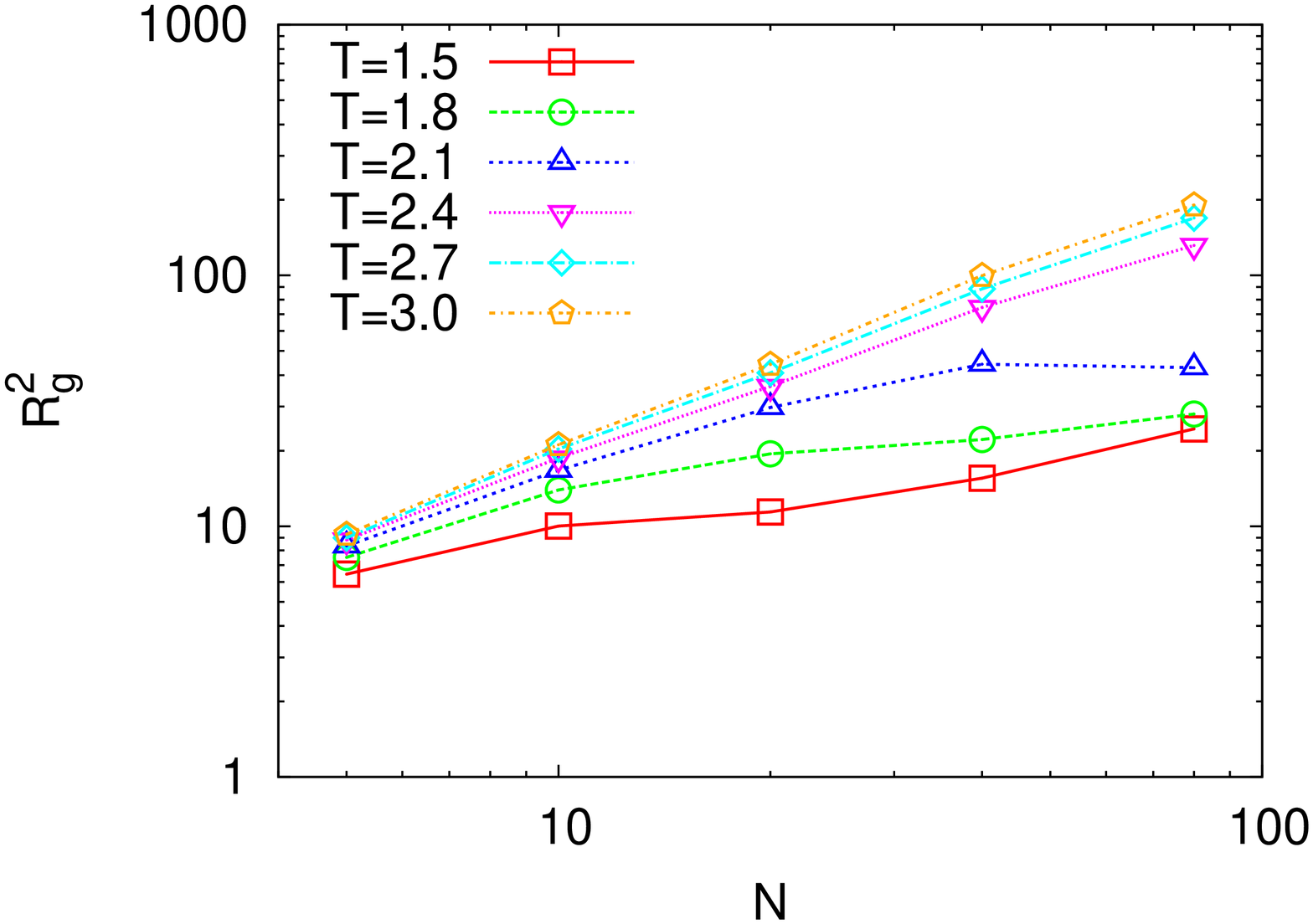}
}}}
\subfloat[][]{
\rotatebox{270}{\resizebox{!}{1.00\columnwidth}{%
   \includegraphics{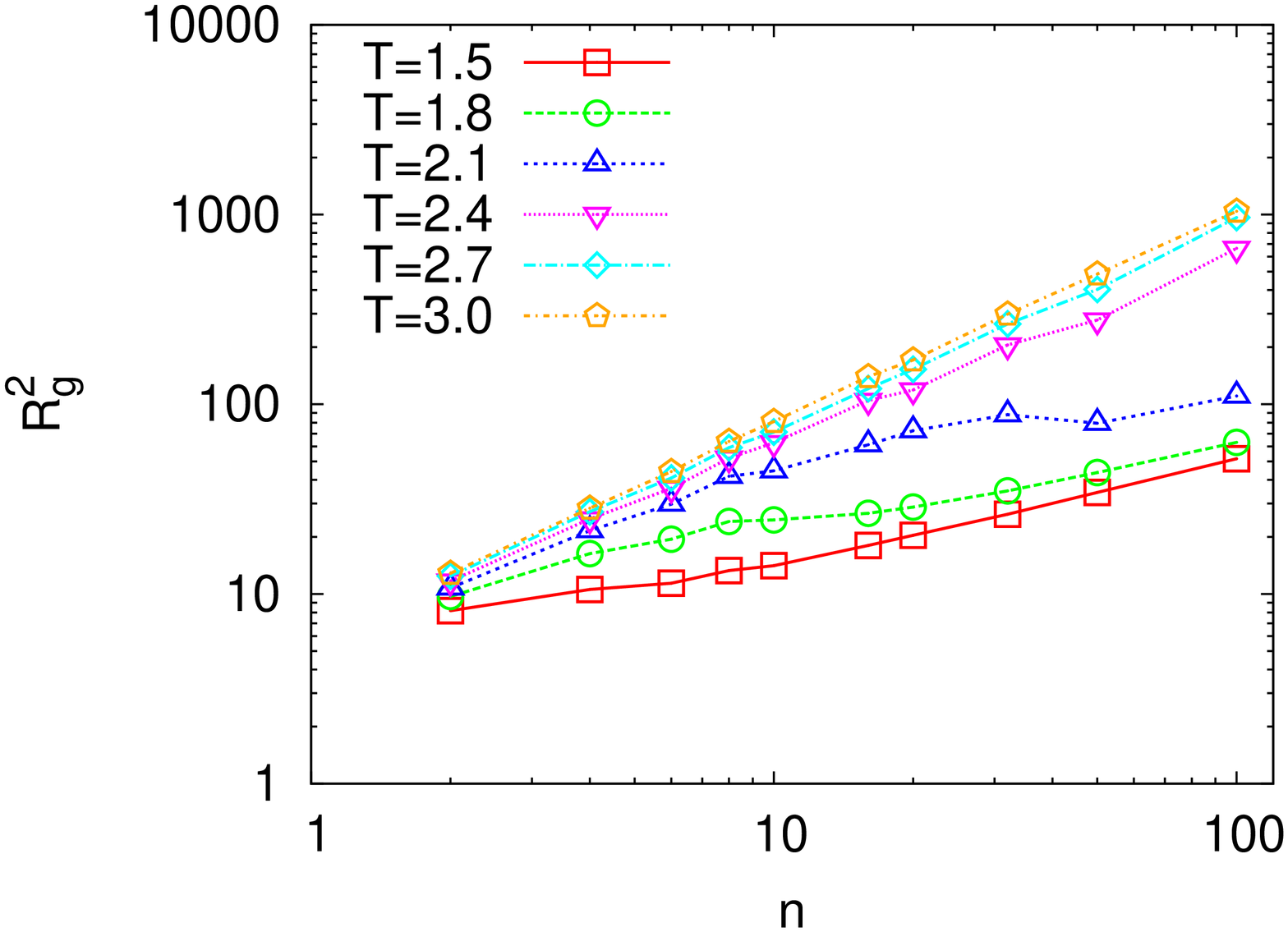}
}}}
\end{center}
\caption{\label{fig11}(Color online)
Two representative cases for the mean square gyration
radius for chains of $n=10$ as a function
of the block length $N$ (a), and for chains of $N=20$ as
a function of the number of blocks $n$ (b).
}
\end{figure*}

Plots showing the overall asphericity of multiblock chains as a
function of $N$ or $n$ (Fig.~\ref{fig5}) neither do reveal any
entry to the phase separation regime of multiblock chains and the
behavior is similar to what is seen for the average properties of
the individual blocks. Asphericity shows an independence on $n$
and $N$ for temperatures close to $\Theta$ (i.e. $T=3.0$), while
$b/S$ decays with $N$ or $n$ obtaining a more globular formation
with decreasing temperature and increasing $n$ or $N$, indicating
the entry to the regime where the multiblock chain collapses in a
poor solvent. Showing the corresponding plots of Fig.~\ref{fig4}
for the asphericity of the blocks (Fig.~\ref{fig6}), we draw the
same conclusions. However, the asphericity of the blocks shows a
more pronounced relation to the entry of the globule regime.
Higher block lengths exhibit a higher dependence on the change of
temperature and decreasing the temperature from $T=2.4$ to $T=1.5$
(Fig.~\ref{fig6}a) the slope of the curve does not change,
implying that regarding asphericity of individual blocks there is
a stronger dependence on $N$, rather than $n$ (Figs.~\ref{fig6}b).

As in the case of Fig.~\ref{fig3} where we have illustrated the
asphericity of a multiblock chain as a function of temperature,
the phase behavior of multiblock copolymer chains can be perfectly
reflected to other properties, as for example the orientational
correlation function of Eq.~(\ref{eq:5}). Then Fig.~\ref{fig7}
shows examples of this quantity for different cases. For a system
not exhibiting full phase separation this orientational
correlation (not shown here) for high values of distance $s$
(defined in number of beads along the chain) is usually neglected,
since in this range of $s$ we rather observe random scattered
points indicating a loss of correlation. However, as we will see
this range of $s$ could provide very important information in the
case of multiblock copolymer chains, when a single interface
between blocks A and B forms for various $n$ and $N$ at low
temperatures $T$. For $s$ close to zero (Fig. ~\ref{fig7}) one is
not able to fit any exponents, rather there occurs a very fast
decay, which is more pronounced at lower temperatures. It is
remarkable that this initial decay is independent of the varied
parameters; an obvious interpretation is that in the scale of a
few subsequent bonds the chain maintains a high local flexibility.

We focus now on our aim which is to relate the phase behavior to
the bond orientational correlations, which is very well shown in
Figs.~\ref{fig7}a and b. Figure~\ref{fig7}a shows results for a
multiblock copolymer chain with $N=60$. At temperatures $T \leq
2.1$ and $s>60$ we observe peaks which are repeated periodically,
as $s$ increases. We can clearly observe four different peaks,
which correspond to the $(n-2)/2$ consecutive A-B blocks
respectively. The centers of these peaks correspond exactly to an
integer multiple of the total length of an A and B block, which
are joint together one after the other. These peaks appear for the
range of temperatures that a complete phase separation of A and B
blocks takes place (a single interface between A and B blocks is
formed). Then the values at small $s$ ($s<2N$) correspond to
correlations of bonds within the length of a block A and B. Again,
the orientational correlation function is able to provide very
good information for intermediate and high $s$ for the phase
behavior of multiblock copolymer chains under poor solvent
conditions. Figure~\ref{fig7}b shows results for the case $n=16$
at $T=1.8$ for chains of different block length $N$. At this
temperature, systems that exhibit a full phase separation between
blocks A and B show symmetrical peaks at positions $2lN$, where
$l$ takes values $2,3,\ldots,(n-2)/2$.

The number of contacts of beads does not also reveal directly any
correspondence to the phase behavior of multiblock copolymers
(Fig.~\ref{fig8}), although these numbers are a direct result of
the microscopic interactions between monomers. The number of
contacts A-A (which is equal to the number of contacts B-B due to
the symmetry of our model) decreases monotonically with increasing
temperature as beads come apart, but with a lower rate for
temperatures $T>2.4$, as we can see from Fig.~\ref{fig8}a. We have
found~\cite{10} that this is the temperature that the chain enters
the collapsed state, i.e. effects due to the poor solvent are
becoming very important until the chain obtains a globular formation
at even lower temperatures. For different number of blocks $n$,
this change in the slope sets in at different temperatures. Of
course, for symmetric multiblock copolymer chains of the same
total length $nN=600$ and increasing $n$ it is expected that the
number of contacts A-A will become lower. Increasing the number of
blocks and joining them one after the other, the number of
contacts A-B increases. This is better shown in Fig.~\ref{fig8}b.
Counting the number of contacts between beads irrespective of
their type (Fig.~\ref{fig8}c), we can see that the increasing
number of blocks $n$ for chains of the same total length $nN$
results in smaller amount of contacts between beads. This is
obviously attributed to unfavorable contacts A-B; beads A and B
prefer to be apart, reducing in this way the number of contacts
$n_{tot}$ for chains composed of the same amount of beads A and B.
Again, such plots do not provide a clear information on the regime
that this phase separation between blocks in a multiblock
copolymer chain takes place. We rather see a change in the
slope of the curves upon entering the
collapsed state, and the results of Fig.~\ref{fig8}
suggest that any transitions take place gradually.

As the bond orientational correlation has shown a direct
correlation on the phase behavior of a multiblock chain, we have
also looked at other properties as a function of $n$, $N$, and
$T$, as it is for example the dihedral angle $D$ which is formed
by four consecutive monomers and taking the average of this
quantity along the chain for all $nN-3$ tetrads disregarding in
this way effects attributed to different sequences along the
chain. As the orientational parameters for small correlation
distances $s$ in terms of monomer count (e.g. $s<4$) has shown,
the chain keeps its local flexibility irrespective of the
interactions A,B, or even the chain architecture~\cite{34}.
Therefore, small differences in $D$ are expected with variation of
the parameters $n$, $N$, and $T$ in the case of multiblock
copolymers, as it also appears in Fig.~\ref{fig9}. Then,
Fig.~\ref{fig9}a illustrates the behavior of $D$ with temperature
$T$. Increase of the temperature induces a slight increase in $D$
for all chains with different number of blocks, that is a tiny
local stretching of the chains. The increase in the number of
blocks $n$ shows that $D$ also increases, but this change is
rather small due to reasons that have already been discussed. In
this case all chains have the same total length $nN=600$. If we
choose a rather intermediate block length (as for example in
Fig.~\ref{fig9}b, $N=40$), we can see the dependence on $n$.
However, in this case and for $n>20$, $D$ has reached a plateau.
For small number of $n$ there is a sharper increase of $D$ with
$n$, which is also more pronounced at lower temperatures and in
the range $n<10$. But, this is not the whole story, $D$ depends
also on the block length $N$ (see Fig.~\ref{fig9}c). Here we show
an example for $n=20$, which corresponds to a point belonging to
the plateau regime of part (c) of Fig.~\ref{fig9}. $D$ provides
interesting information on the local behavior of multiblock
copolymer chains, with small differences arising by variation of
the parameters and no effects related to the phase behavior of the
multiblock copolymers can be found, as expected, since $D$ is a
particularly local property of the chain.

Figure~\ref{fig10}a shows the dependence on $N$ for a choice of
intermediate number of blocks $n=20$ for different temperatures.
Differences due to temperature are expected to increase as the
block length $N$ increases. We see that small differences result
from the variation of $N$ for different temperatures. Similar
behavior we see for other values of $n$. Interestingly the size of
individual blocks also depends on the number of blocks $n$
(Fig.~\ref{fig10}b) at lower temperatures, but at higher
temperatures the gyration radius of the blocks does not change
with $n$. Again, these are results for an intermediate value of
block length $N=40$, where a full phase separation between A and B
blocks can take place at low temperatures. One could argue that
the number of blocks does not play any role when the chain has a
coil formation, while in the collapsed state a $n$-dependence
occurs at low number of blocks $n$. However, a connection of phase
separation or a partial phase separation between blocks is not
seen for this property of individual blocks by examining all
relevant cases and possible variations of parameters.
Nevertheless, it shows us that the dependence on $n$ of the
individual size of the blocks seen under poor solvent conditions,
is smeared out at high number of blocks $n$. Similar effects are
seen for other choices of $N$. Unfortunately, increasing the
number of blocks $n$ for the interesting range of $N$ and apply
further exploration of this phenomena is rather limited in the
present simulations.

Assuming a rather small number of blocks $n$, and knowing that
full phase separation is favored by a small number of blocks, we
plot the squared average gyration radius of the whole multiblock
chain (Fig.~\ref{fig11}a). At temperatures close to $\Theta$ one
can see a straight line (mind the double logarithmic scale), as
expected. Deviations from this behavior are seen for lower
temperatures. The dependence on the number of blocks $n$ is shown
in Fig.~\ref{fig11}b. In both figures we can rather see the same
behavior for variation of $n$ and $N$, showing that the overall
dimensions of the multiblock copolymer chain rather do not depend
very much on the way that the blocks are distributed along the
chains. Of course, increasing the number of blocks favors a more
elongated chain at lower temperatures, due to the higher number of
unfavorable interactions between A and B monomers. Similar
conclusions are drawn by examining all combinations of $n$, $N$,
and $T$.

\section{SUMMARY}
\label{sec:4}

In this manuscript we have presented results for a single
multiblock chain and its individual blocks varying parameters such
as the block length $N$, the number of blocks $n$, and the
temperature $T$. The phase behavior of multiblock copolymer chains
in a dilute solution has been discussed in detail
elsewhere~\cite{11,12}, as well as the static properties of the
clusters composed of blocks A or B~\cite{10}, which are formed due
to the microphase separation of monomers A and B. In this
manuscript we have tried to make a connection between the phase
behavior of these macromolecules and the static properties of the
chain and of the individual blocks on average. We have shown that
properties related to the chain as a whole can reveal significant
information on the phase behavior of multiblock copolymers, while
study of the individual blocks does not provide such information.
The presence of a single A-B interface between blocks of different
type play the major role in the behavior of multiblock chains. The
effect of the presence of this interface is reflected to
properties such as the bond orientation correlation function and
the asphericity of the whole multiblock chains. In particular, we
have shown that correlations for intermediate and high distances
$s$ (measured in number of monomers) along the chains for the
orientation correlation function which, is generally not
interesting for a homopolymer chain, can show very interesting
effects due to the presence of a well defined A-B interface in the
case of multiblock copolymers. Furthermore, it is remarkable that
the study of properties of the individual blocks  can reveal
dependence on parameters such as the number of blocks $n$ as the
chain gets into the collapsed state. This dependence is lost for
higher number of blocks. Phase separation that is due to the
microscopic interactions between monomers and the incompatibility
of A and B monomers can be mainly measured with properties related
to the chain as a whole, or on the basis of a cluster analysis for
multiblock copolymer chains, as was done
previously~\cite{10,11,12}.

The results presented in this manuscript, in combination with
recent results in the phase behavior of multiblock copolymer
chains, give a detailed description for these systems, at least
for the most symmetrical case, where an even number of blocks of A
and B type monomers alternate along a linear multiblock chain with
$N_{A}=N_{B}$ being the lengths of blocks A and B respectively. It
is interesting to see that the change of composition of such
symmetrical chains can lead to different scenarios of phase
separation. Introducing any kind of asymmetry in our systems or
the consideration of an additional type of monomers in the
structure of multiblock chains could result in different and
interesting behavior for these systems. The use of a standard
generic coarse-grained model to describe our chains could act as a
reference system for the study of these complicated systems.
Moreover, the behavior of multiblock copolymers can be
parallelized to that of various biological macromolecules which
are formed by periodically repeated chemical units (``monomers'')
along their chain. A detailed discussion on the properties of
multiblock copolymer chains using a simple model for a variety of
parameters $n$, $N$, and $T$ is useful to understand the behavior
of these chains, before any other effects, due to, e.g. charges,
more complicated interactions, architecture, etc., come into play.
We envisage that the present contribution could also stimulate
further work on these topics.

\begin{acknowledgements}

P.E.T. is grateful for financial support by the Austrian Science
Foundation within the SFB ViCoM (grant F41). He also acknowledges
financial support through a Max Planck fellowship awarded by the
Max Planck Institute for Polymer Research at an initial stage of
this work.

\end{acknowledgements}

\end{document}